\documentclass[aps,prl,twocolumn,nofootinbib,showpacs,floatfix,superscriptaddress,preprintnumbers]{revtex4}
\usepackage{amsmath}
\usepackage{amssymb}
\usepackage{epsfig}
\usepackage{graphicx}
\usepackage{rotating}
\usepackage{pstricks}
\usepackage{pst-coil}
\DeclareMathAlphabet{\mathsc}{OT1}{cmr}{m}{sc}
\newcommand {\ignore}[1]{}

\usepackage{color} 
\usepackage{multirow} 
\usepackage[T1]{fontenc} 
\usepackage{slashed}

\def\10{$SO(10)$}
\def\21{SU(2) $\otimes$ U(1) }

\def\422{$SU(4) \otimes SU(2) \otimes SU(2)$}
\def\321{SU(3) $\otimes$ SU(2) $\otimes$ U(1)}
\def\gsim{\raise0.3ex\hbox{$\;>$\kern-0.75em\raise-1.1ex\hbox{$\sim\;$}}}
\def\lsim{\raise0.3ex\hbox{$\;<$\kern-0.75em\raise-1.1ex\hbox{$\sim\;$}}}

\def\lsim{\raise0.3ex\hbox{$\;<$\kern-0.75em\raise-1.1ex\hbox{$\sim\;$}}}
\def\gsim{\raise0.3ex\hbox{$\;>$\kern-0.75em\raise-1.1ex\hbox{$\sim\;$}}}

\def \znbb {0\nu\beta\beta}

\newcommand{\ba}{\begin{array}}
\newcommand{\ea}{\end{array}}
\relax

\def\321{$SU(3)\times SU(2)\times U(1)$}

\newcommand{\AddrAHEP}{
  {\it AHEP Group, Instituto de F\'{\i}sica Corpuscular --
    C.S.I.C./Universitat de Val{\`e}ncia \\
    Edificio de Institutos de Paterna, Apartado 22085,
  E--46071 Val{\`e}ncia, Spain}}

\def\gsim{\raise0.3ex\hbox{$\;>$\kern-0.75em\raise-1.1ex\hbox{$\sim\;$}}}
\def\lsim{\raise0.3ex\hbox{$\;<$\kern-0.75em\raise-1.1ex\hbox{$\sim\;$}}}

\begin{document}

\preprint{IFIC/12-42}

\renewcommand{\Huge}{\Large}
\renewcommand{\LARGE}{\Large}
\renewcommand{\Large}{\large}
\def \znbb {$0\nu\beta\beta$ }
\def \nbb {$\beta\beta_{0\nu}$ }
\title{Bi-large neutrino mixing and the Cabibbo angle}  \date{\today}
\author{S. M. Boucenna} \email{boucenna@ific.uv.es} \affiliation{\AddrAHEP}
\author{S. Morisi} \email{morisi@ific.uv.es} \affiliation{\AddrAHEP}
\author{M. T{\'o}rtola}\email{mariam@ific.uv.es}    \affiliation{\AddrAHEP}
\author{J.~W.~F.~Valle} \email{valle@ific.uv.es} \affiliation{\AddrAHEP}
\date{\today}

\begin{abstract}

  Recent measurements of the neutrino mixing angles cast doubt on the
  validity of the so-far popular tri-bimaximal mixing ansatz.  We
  propose a parametrization for the neutrino mixing matrix where the
  reactor angle seeds the large solar and atmospheric mixing angles,
  equal to each other in first approximation.  We suggest such {\it
    bi-large} mixing pattern as a model building standard, realized
  when the leading order value of $\theta_{13}$ equals the Cabibbo
  angle $\lambda_C$.

\end{abstract}


\maketitle

\section{introduction}

Understanding the structure of neutrino mixing from first principles
is part of the flavor problem of the Standard Model, one of the
deepest in particle physics.
A partial useful strategy is to formulate attractive mixing patterns
that may help to seek for possible underlying flavor symmetries.
The tri-bimaximal mixing (TBM) ansatz~\cite{Harrison:2002er} for
describing the neutrino mixing matrix~\cite{schechter:1980gr} assumes
a maximal atmospheric angle and zero reactor angle, as was suggested
by the experimental data.  Such striking features point towards an
underlying symmetry and indeed, many models based on (mostly discrete)
non-abelian flavour symmetries were successful in reproducing this
ansatz~\cite{babu:2002dz,altarelli:2005yp}.  Small deviations from the
TBM are expected on theoretical grounds.

However the recent results published by the
Double-Chooz~\cite{Abe:2011fz}, Daya Bay~\cite{An:2012eh},
RENO~\cite{Ahn:2012nd}, T2K~\cite{Abe:2011sj} and
MINOS~\cite{nichol-nu2012} collaborations, indicate that the the
reactor angle is relatively large so that corrections to the TBM
pattern should be, in fact, quite large, casting doubt on its validity
as a good first approximation reproducing the neutrino mixing pattern.
To be more precise, on theoretical grounds a small deviation of order
of the square of the Cabibbo angle was expected for the reactor angle,
while recent observations indicate a much larger value of about the
order of the Cabibbo angle.
To evade this problem, different ansatz have been considered like the
bimaximal mixing \cite{Altarelli:2009gn} or the Golden ratio, see
\cite{Albright:2010ap} for a review.  However, most of these models
assume a $\mu-\tau$-invariant structure in order to predict a maximal
atmospheric mixing angle.  On the other hand, at the Neutrino 2012
conference the MINOS collaboration also gave hints for a non-maximal
atmospheric mixing angle.

\vskip2.mm
Here we propose a different approach where we take the reactor mixing
angle as the fundamental parameter. As will be shown below, the
resulting parametrization does not reproduce the TBM pattern as a
limiting case, though a maximal atmospheric angle can be obtained.
The main idea is that, since the reactor angle is the only small
mixing parameter for the leptons, we can use it to seed both the solar
and atmospheric mixing angles, as follows,
\begin{equation}\label{s13}
\begin{array}{l}
\sin\theta_{13}=\lambda\,;\\
\sin\theta_{12}=s\,\lambda\,;\\
\sin\theta_{23}=a\,\lambda\,,
\end{array}
\end{equation}
where the small parameter $\lambda$ is the reactor angle, while 
$s\,\simeq\,a$ are free parameters of order few. Solar and atmospheric
mixings are expressed in terms of a linear dependence on the reactor
angle.
In the limit where $\lambda \to 0 $ neutrinos are unmixed.  

Using the general symmetric parametrization of the neutrino mixing
matrix~\cite{schechter:1980gr} one can trivially obtain a simple
approximate description by expanding only in the small parameter
$\lambda$.
For example, the Jarlskog-like invariant describing CP violation in
neutrino oscillations is then given by:
\begin{equation}\label{Jcp}
J_{CP} \approx a~s~\lambda^{3}\sqrt{1-a^2 \lambda^2} \sqrt{1-s^2 \lambda^2} \sin(\phi_{13}-\phi_{12}-\phi_{23}) 
\end{equation}
given explicitly in terms of the rephase-invariant Dirac combination.
Likewise, the effective mass parameter describing the amplitude for
neutrinoless double-beta decay is given in terms of the two Majorana
CP phases.  

In what follows, for simplicity, we take all parameters to be real.
In order to fix the values of the large parameters $s$ and $a$ we
consider the latest experimental results on neutrino oscillation
parameters.  At the Neutrino 2012 conference in Kyoto the MINOS
Collaboration has reported a non-maximal value for the atmospheric
mixing angle~\cite{nichol-nu2012}:

\begin{equation}
\label{eq:23m}
  \sin^22\theta_{23}=0.94^{+0.04}_{-0.05}\,\footnote{In a
three-neutrino analysis
    this quantity will be equal to $4 |U_{\mu 3}|^2 (1- |U_{\mu 3}|^2)
= 4 \cos^2\theta_{13}\sin^2\theta_{23}
(1-\cos^2\theta_{13}\sin^2\theta_{23})$.}
\end{equation}
with maximal mixing disfavoured at 88\% C.L. This result comes from
the analysis of $\nu_\mu$ disappearance in the MINOS accelerator beam
and corresponds to two degenerate points for $\sin^2\theta_{23}$,
namely
\begin{equation}
\sin^2\theta_{23} = 0.38~~~~~\mathrm{and}~~~~~
\sin^2\theta_{23} = 0.62.
\end{equation}

This comes from the fact that the disappearance channel is
octant-symmetric and therefore MINOS data by themselves can not show a
preference for a given octant of $\theta_{23}$. However, one may
expect that in combination with the searches for electron-neutrino
appearance at the long-baseline experiments MINOS and T2K, and with
the recent measurements of $\theta_{13}$ at reactor experiments the
degeneracy in Eq.~(\ref{eq:23m}) will be broken and one octant will be
preferred over the other.
Global analysis so far have not been able to give a completely clear
picture about the true octant of $\theta_{23}$. The analysis given
in~\cite{Tortola:2012te} indicates small deviations of maximality,
with the octant preference correlated with the mass hierarchy. This
correlation is also seen in the recent analysis of atmospheric
neutrino data in Super-Kamiokande~\cite{itow-nu2012}.
However, the analyses given
in~\cite{Fogli:2012ua,GonzalezGarcia:2010er} have shown a preference
for $\theta_{23} < \pi/4$ with different levels of significance.
All previous neutrino oscillation global fits agree in getting
$\theta_{23}$ in the first octant for normal mass hierarchy, so for
the purpose of this article we will assume that this is the case.

\section{bi-large mixing}

Here we discuss the recent neutrino oscillation results in terms of
our proposed ansatz.  As already pointed out above, the recent
experimental data provide a robust measurement of a relatively large
$\theta_{13}$. 
  
  Using the best fit values of the mixing angles in
  Ref.\,\cite{Tortola:2012te} or Ref.\,\cite{Fogli:2012ua}, we can
  fix the three parameters in Eq.\,(\ref{s13}): $\lambda\sim 0.16\,\,
  (0.15)$, $a\sim 4.13 \,\,(4.21)$ and $s\sim 3.53 \,\,(3.65)$.
Then, from the data we can directly read that
\begin{equation}
\quad \sin\theta_{12}~=~\mathcal{O}(\sin\theta_{23})\,\,,
\end{equation}

Now we go a step further and assume the following
\begin{equation}
\quad \sin\theta_{12}~=~\sin\theta_{23}\,\,,
\end{equation}
which in our parametrization means
\begin{equation}
\label{eq:s=a}
s\,=\,a.
\end{equation}
Since both solar and atmospheric angles are large we call this case
{\it bi-large mixing ansatz}.

\vskip5.mm
Suppose now that we are given some model predicting bi-large mixing
$a\,=\,s$ at leading order.  Next-to-leading order operators in the
Lagrangian in general induce deviations from the reference values in
Eq.~(\ref{s13}) which may be reliably determined within a given model.
Here we present a simple model-independent estimate of such
corrections, obtained as follows.
Typically it is expected that the corrections to the three mixing
angles from next to leading order terms are of the same order, that is
$\sin\theta_{ij}\to\sin\theta_{ij}\pm \epsilon$ where we have
introduced a new parameter $\epsilon$ to characterize the magnitude of
the correction.
In this case our bi-large mixing gets corrections of the same order
for the three mixing angles (given by $\epsilon$) and which may either
increase or decrease the starting bi-large values of the mixing
angles.  For definiteness let us consider an example where bi-large
mixing is corrected as
\begin{equation}\label{eq:sa}
\begin{array}{l}
\sin \theta_{13}=\lambda-\epsilon \,;\\
\sin\theta_{12}=s \lambda-\epsilon \,;\\
\sin\theta_{23}=a \lambda+\epsilon \,.
\end{array}
\end{equation}
where we take $s=a$ as in Eq.~(\ref{eq:s=a}).  Since we have three
free parameters, we can fix them using the best fit values reported by
global analysis of neutrino oscillation data in
Refs.~\cite{Tortola:2012te,Fogli:2012ua}. The results are given in
Table\,\ref{tab1}.
\begin{table}[h!]
\begin{center}
\begin{tabular}{|l|c|c|c|}
\hline
Ref. & $\lambda$ & $s$  & $\epsilon$ \\
\hline
Forero {\it et al.} \cite{Tortola:2012te} & $0.23\pm0.04$ & $2.8^{+0.5}_{-0.4}$ & $0.067^{+0.035}_{-0.025}$ \\
\hline
Fogli {\it et al.} \cite{Fogli:2012ua} & $0.19^{+0.03}_{-0.02}$ & $3.0^{+0.5}_{-0.3}$ &
$0.038^{+0.019}_{-0.018}$ \\
\hline
\end{tabular}\caption{Best fit values and $1\sigma$ ranges for the
  parameters $\lambda$, $s$ and $\epsilon$ in Eq.~(\ref{eq:sa}) 
  according to the global fits to
  neutrino oscillations.}\label{tab1}
\end{center}
\end{table}

\begin{figure}[h!]
\begin{center}
 \includegraphics[width=7cm]{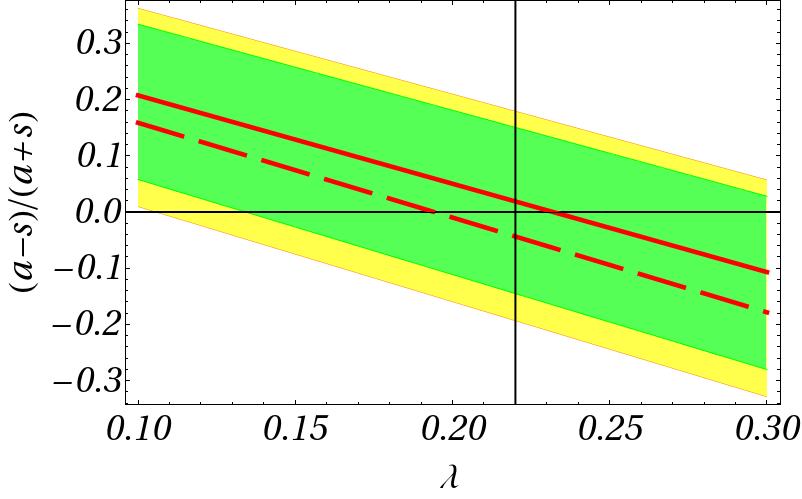}
 \caption{ Deviation from the bi-large ansatz versus the expansion
   parameter $\lambda$ at two and three sigma in the neutrino
   oscillation parameters. The solid and dashed lines
  indicate the best fits of Refs.~\cite{Tortola:2012te} and
\cite{Fogli:2012ua}, respectively.  The strict bi-large
   ansatz holds when $\lambda \simeq \lambda_C$ (vertical line). }
\label{fig1}\end{center}
\end{figure}

In order to quantitatively clarify the role of the relation in
Eq.\,(\ref{eq:s=a}) with respect to the reactor mixing angle, we
consider here the most generical case given by Eq.~(\ref{eq:sa})
where the three angles are given in terms of four parameters instead
of three. Three of these parameters can be fixed from the three
measured mixing angles, leaving one free parameter that we choose to
be $\lambda$. In order to quantify the deviation from our exact
bi-large mixing ansatz defined in Eq.\,(\ref{eq:s=a}) we plot the
combination $(a-s)/(a+s)$ as a function of the expansion parameter
$\lambda$ in Fig.~\ref{fig1}.  The colored/shaded bands are calculated
from the two and three sigma allowed ranges for the neutrino
oscillation parameters obtained in the current global fits. The solid
and dashed lines indicate the best fits of Refs.~\cite{Tortola:2012te}
and \cite{Fogli:2012ua}, respectively.
It is remarkable that the strict bi-large ansatz in Eq.~(\ref{eq:s=a})
holds when $\lambda \simeq \lambda_C$ where $\lambda_C\approx
0.22$. This means that $\lambda_C$ is the leading order value of
$\sin\theta_{13}$.

\vskip5.mm

\begin{figure}[h!]
\begin{center}
 \includegraphics[width=7cm]{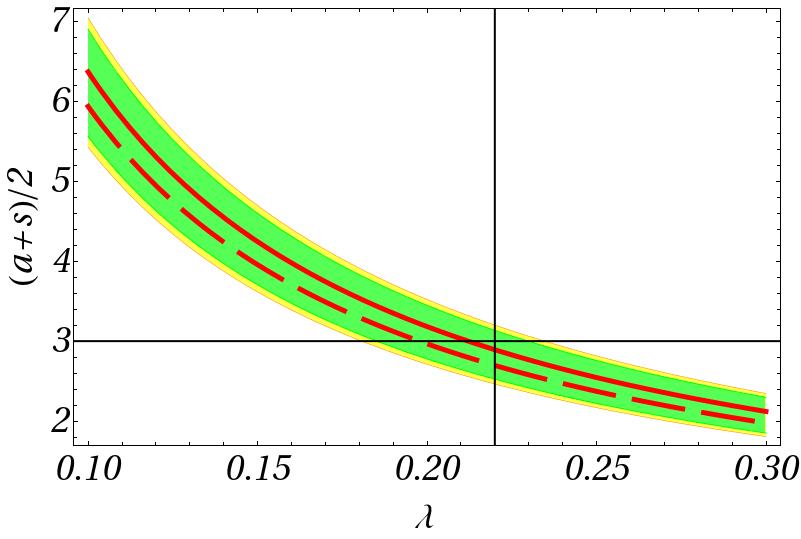}
 \caption{Average of solar and atmospheric angles versus the expansion
   parameter $\lambda$ at two and three sigma in the neutrino
   oscillation parameters. The solid and dashed lines
indicate the best fits of Refs.~\cite{Tortola:2012te} and
\cite{Fogli:2012ua}, respectively.}
\label{fig2}\end{center}
\end{figure}

In Fig.~\ref{fig2} we show the average value of $a$ and $s$, that is
$(a+s)/2$, as a function of $\lambda$. As displayed in the figure, the
correlation is such that $(a+s)/2 \sim 3$ when $\lambda\sim\lambda_C$.
It follows that one possible form of our bi-large ansatz, which can be
useful for model building, is
\begin{equation}\label{a3}
\begin{array}{l}
\sin \theta_{13}=\lambda;\\
\sin\theta_{12}=3\,\lambda;\\
\sin\theta_{23}=3\,\lambda\,.
\end{array}
\end{equation}
It is remarkable to see how such a simple form is nearly consistent
with current global neutrino oscillation data.

Now a few words on model building. 
Given a particular mixing matrix $U$, the structure of the neutrino
mass matrix is fixed as $$m_\nu = U \cdot D \cdot U^T$$ where $D$ is a
diagonal matrix. For the sake of illustration, we consider in our
mixing ansatz in Eq.~(\ref{a3}) a normal and strongly hierarchical
spectrum ($m_{\nu_1} =0$) for neutrino masses and fix the square mass
differences at their best fit values, as given in
Ref.\,\cite{Tortola:2012te}. We find that the resulting weak-basis
neutrino mass matrix $m_\nu$ has the form
\begin{eqnarray}
m_\nu\sim
\left(
\begin{array}{ccc}
0.20 & 0.32 & 0.15 \\
         & 0.75 & 0.70 \\
         &         & 1
\end{array}
\right)\sim
\left(
\begin{array}{ccc}
\lambda_C & \lambda_C & \lambda_C \\
         & 1 & 1 \\
         &         & 1
\end{array}
\right)\,
\label{eq:fn1}
\end{eqnarray}
where in the last step the Cabibbo angle is used as the expansion
parameter and we do not specify numerical coefficients of order one.

From the form obtained in Eq.~(\ref{eq:fn1}) one sees that the
parameter $\lambda_C$ appears only in the first row.
On the other hand in the ``atmospheric sector'' there seems to be
``democracy'' in the choice of the neutrino mass entries. Altogether
the above indicates two general features regarding the neutrino mass
generation mechanism.
\begin{itemize}
\item a Frogatt-Nielsen-like flavour symmetry~\cite{Froggatt:1978nt}
  that could perhaps generate the required pattern given in
  Eq.~(\ref{eq:fn1}).
\item some gravity-like~\cite{degouvea:2000jp} ``flavor-blind''
  mechanism operating within the ``atmospheric sector''. The problem
  in this case is that, for reasonable values of the coefficients of
  the dimension five operator, the induced neutrino masses are too
  small.  However there may be other well-motivated ``anarchy''-type
  schemes~\cite{Hall:1999sn,gouvea-pascos-2012}.
\end{itemize}

It is beyond the scope of this paper to provide a definite model
realization, but rather to stress the simplicity of the ansatz in
Eq.~(\ref{eq:fn1}) which may provide a fresh model-building guideline
that may perhaps replace the tri-bimaximal ansatz.

\section{summary and outlook}
\label{sec:summary-outlook}

We have argued that recent experimental results on neutrino
oscillations are very well-described by the ansatz in Eq.~(\ref{s13})
with small deviations, as in Eqs.~(\ref{eq:sa}). It is truly
remarkable that with the expansion parameter $\lambda$ taken as
$\lambda \simeq \lambda_C$, the Cabibbo angle characterizing the quark
mixing matrix, we obtain the simplest bi-large limit, $s = a$, given
in Eq.~(\ref{eq:s=a}). This appears to be an important numerological
``coincidence'' which may drastically change our theoretical approach
for constructing neutrino mass models, by moving from a {\it
  geometrical} interpretation of the neutrino mixing angles to one in
which these are no longer associated to Clebsh-Gordan coefficients of
any symmetry, in sharp contrast to the previous paradigm~\footnote{We
  should stress, however, that the TBM pattern may still be tenable if
  the underlying theory is capable of providing sufficiently large
  corrections to $\theta_{13}$ without affecting too much the solar
  angle, which is in principle possible.}.
The simple correlations between ansatz deviations and $\lambda$ are
illustrated in Figs.~\ref{fig1} and \ref{fig2}. The bi-large ansatz
holds for $\lambda \simeq \lambda_C$ (vertical line).  It remains to
be seen whether nature is perhaps telling us something profound
regarding the ultimate theory of flavour.

\section*{Acknowledgments}

We thank Martin Hirsch for discussions. This work was supported by the
Spanish MINECO under grants FPA2011-22975 and MULTIDARK CSD2009-00064
(Consolider-Ingenio 2010 Programme), by Prometeo/2009/091 (Generalitat
Valenciana), by the EU ITN UNILHC PITN-GA-2009-237920. M.T.\
acknowledges financial support from CSIC under the JAE-Doc programme,
co-funded by the European Social Fund.
S.M. acknowledges Juan de la Cierva contract.


\begin{thebibliography}{10}

\bibitem{Harrison:2002er}
P.~F. Harrison, D.~H. Perkins and W.~G. Scott,
\newblock Phys. Lett. {\bf B530}, 167 (2002).

\bibitem{schechter:1980gr}
J.~Schechter and J.~W.~F. Valle,
\newblock Phys. Rev. {\bf D22}, 2227 (1980);
W.~Rodejohann and J.~W.~F. Valle,
\newblock Phys.Rev. {\bf D84}, 073011 (2011), [1108.3484].

\bibitem{babu:2002dz}
K.~S. Babu, E.~Ma and J.~W.~F. Valle,
\newblock Phys. Lett. {\bf B552}, 207 (2003), [hep-ph/0206292].

\bibitem{altarelli:2005yp}
G.~Altarelli and F.~Feruglio,
\newblock Nucl. Phys. {\bf B720}, 64 (2005), [hep-ph/0504165].

\bibitem{Abe:2011fz}
DOUBLE-CHOOZ Collaboration, Y.~Abe {\em et~al.},
\newblock Phys.Rev.Lett. {\bf 108}, 131801 (2012), [1112.6353].

\bibitem{An:2012eh}
DAYA-BAY Collaboration, F.~An {\em et~al.},
\newblock Phys.Rev.Lett. {\bf 108}, 171803 (2012), [1203.1669].

\bibitem{Ahn:2012nd}
RENO collaboration, J.~Ahn {\em et~al.},
\newblock 1204.0626.

\bibitem{Abe:2011sj}
T2K Collaboration, K.~Abe {\em et~al.},
\newblock Phys.Rev.Lett. {\bf 107}, 041801 (2011), [1106.2822].

\bibitem{nichol-nu2012}
R.~Nichol,
\newblock Plenary talk at the Neutrino 2012 conference, http://neu2012.kek.jp/.

\bibitem{Altarelli:2009gn} 
  G.~Altarelli, F.~Feruglio and L.~Merlo,
\newblock  JHEP {\bf 0905}, 020 (2009), [0903.1940].


\bibitem{Albright:2010ap}
C.~H. Albright, A.~Dueck and W.~Rodejohann,
\newblock Eur.Phys.J. {\bf C70}, 1099 (2010), [1004.2798].

\bibitem{Tortola:2012te} D.~Forero, M.~Tortola, J.~W.~F.~Valle,
  \newblock arXiv:1205.4018, this updates previous results in New
  J. Phys. 13 (2011) 109401 and New J.Phys. 13 (2011) 063004

\bibitem{itow-nu2012}
Y.~Itow,
\newblock Plenary talk at the Neutrino 2012 conference, http://neu2012.kek.jp/.

\bibitem{Fogli:2012ua}
G.~Fogli {\em et~al.},
\newblock arXiv:1205.5254.

\bibitem{GonzalezGarcia:2010er}
M.~Gonzalez-Garcia, M.~Maltoni and J.~Salvado,
\newblock JHEP {\bf 1004}, 056 (2010), [1001.4524].

\bibitem{Froggatt:1978nt}
C.~D. Froggatt and H.~B. Nielsen,
\newblock Nucl. Phys. {\bf B147}, 277 (1979).

\bibitem{degouvea:2000jp}
A.~de~Gouvea and J.~W.~F. Valle,
\newblock Phys. Lett. {\bf B501}, 115 (2001), [hep-ph/0010299].

\bibitem{Hall:1999sn}
L.~J. Hall, H.~Murayama and N.~Weiner,
\newblock Phys.Rev.Lett. {\bf 84}, 2572 (2000), [hep-ph/9911341].

\bibitem{gouvea-pascos-2012}
A.~Gouvea,
\newblock Plenary talk at the PASCOS 2012 conference, http://neu2012.kek.jp/.

\end{thebibliography}
\end{document}